\documentclass[
twocolumn,
pre,
aps,
superscriptaddress,
showpacs,
preprintnumbers,
amsmath,
amssymb]{revtex4-1}
\usepackage{bm}
\usepackage{color}
\usepackage{graphicx}

\begin{document}

\title {Simplicial complexes and complex systems}

\author{Vsevolod Salnikov}
\affiliation{NaXys, Universite de Namur, 5000 Namur, Belgium}
\author{Daniele Cassese}
\affiliation{NaXys, Universite de Namur, 5000 Namur, Belgium}
\affiliation{Emmanuel College, University of Cambridge, Cambridge, UK}
\author{Renaud Lambiotte}
\email{renaud.lambiotte@unamur.be}
\affiliation{Mathematical Institute, University of Oxford, Oxford, UK}

\begin{abstract}
We provide a short introduction to the field of topological data analysis and discuss its possible relevance for the study of complex systems. Topological data analysis provides a set of tools to characterise the shape of data, in terms of the presence of holes or cavities between the points. The methods, based on notion of simplicial complexes, generalise standard network tools by naturally allowing for many-body  interactions and providing results robust under continuous deformations of the data. We present strengths and weaknesses of current methods, as well as a range of empirical studies relevant to the field of complex systems, before identifying future methodological challenges to help understand the emergence of collective phenomena.
\end{abstract}

\maketitle

\section {Introduction}

Take a cloud of points in a D-dimensional space. The points could correspond to the coordinates of individuals in a space of attributes, such as their age, income and height. Or the positions of birds in the sky. Or the locations of McDonalds in a city. Or a sample of coordinates in phase space for a dynamical system. This type of data is prevalent in a variety of domains, including many areas of Physics, and several mathematical tools have been developed to reveal information hidden in their noisy patterns and to reduce the system dimensionality. Important families of methods include Principal Component Analysis \cite{bishop}, looking for dominant directions to explain the variance in the data, geometric or k-nearest neighbour graphs \cite{barthelemy2011spatial}, where pairs of points are connected if they are sufficiently close, or their fractal dimension \cite{mandelbrot1982fractal} revealing the self-similarity of the data. Each tool reveals a certain aspect of the data, putting emphasis on either statistics, connections or geometry. Topological Data Analysis (TDA) provides an alternative set of tools that extracts  topological features from the data that are invariant under certain transformations, and aims at characterising its shape by means of its number of cavities, holes or voids \cite{muhammad2005graphs, Edelsbrunner08,Otter17}. Topological data analysis can be seen as an extension of the aforementioned families of methods, and intuitively understood in terms of networks with non-binary interactions with a geometrical flavour. 

TDA has gained popularity in the field of data mining and has been applied to data in a broad range of disciplines. The main purpose of this paper is to investigate its potential advantages and limitations for the study of complex systems. To do so, we  first provide a short introductory guide for complex systems scientists to TDA, clarifying underlying concepts, with a particular focus on the theory of simplicial complexes and persistent homology, and pointing to relevant introductory bibliography. We then present salient results of applications of TDA to examples of complex systems. Finally, we try to delineate situations when these tools might be of relevance and identify  steps to turn TDA from a set of computational tools to a framework for a science of complex topology.

\section {Topological data analysis in a nutshell}

\begin{figure}
\centering
\includegraphics[width=0.5\textwidth]{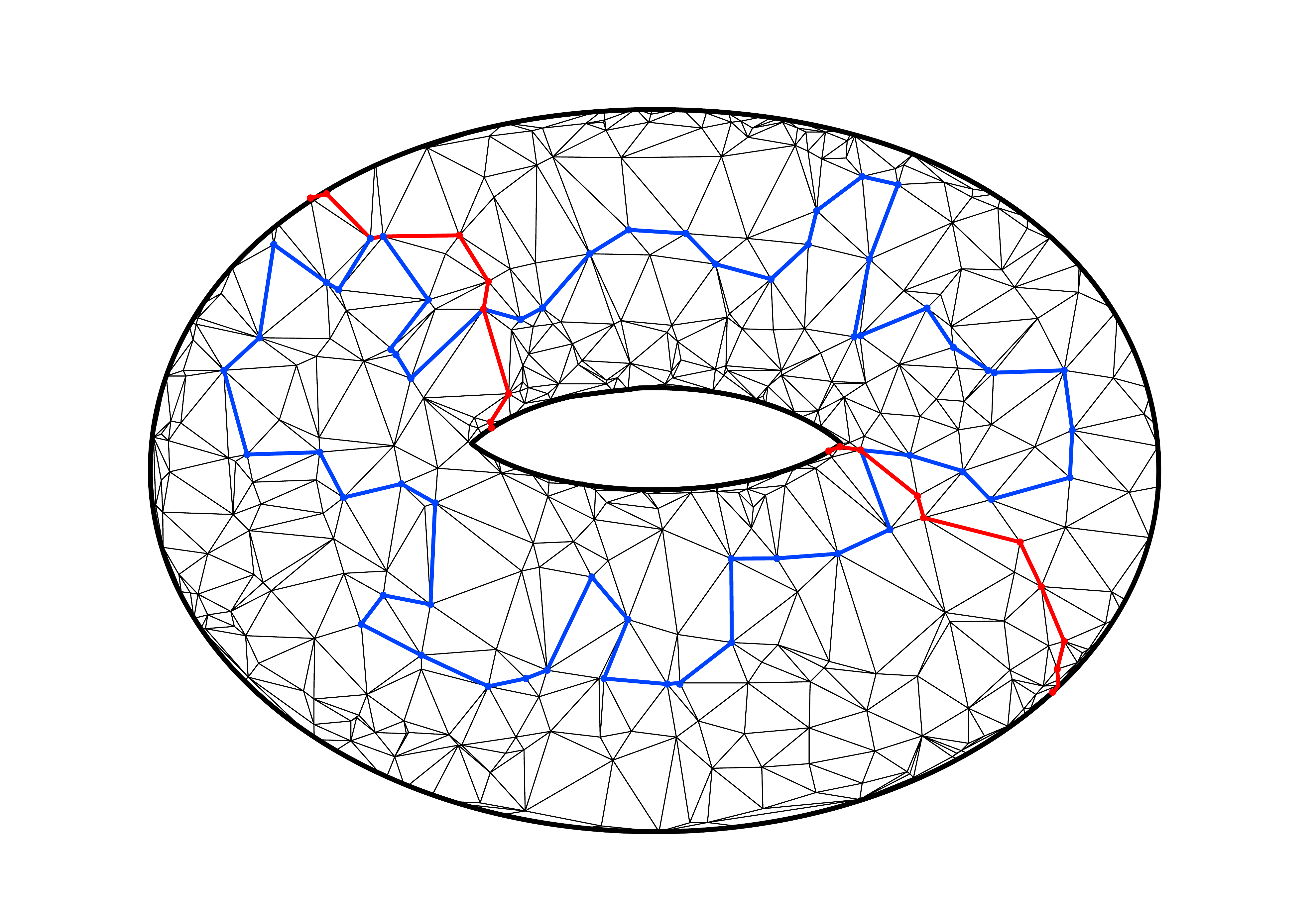}
\caption{Generators of the first homology groups for a triangulation of the torus. The torus has two 1-dimensional holes, the one surrounded by the blue cycle and the one surrounded by the red cycle, so $H_1$ has two elements, which are the equivalence classes of 1-cycles. For example the two red cycles (and any similar cycle on the torus) are in the same equivalence class because they "surround the same hole" while the blue cycle (and any other similar) are in the other equivalence class of $H_1$. The Betti numbers are $\beta_0 =1$ (one connected component), $\beta_1 =2$ (two 1-dimensional holes surrounded by a chain of edges), $\beta_2 =1$ (one void enclosed within the surface).}
\end{figure}

The field of geometry studies properties of an object that are invariant under rigid motion, such as the angles in a triangle or the curvature of a surface. Topology studies instead invariants under continuous deformations, called homotopies, that can be understood as the stretching and shrinking of an object. The larger set of transformations leads to a characterisation of the shape of a topological space. TDA offers computational tools to translate raw data into structured sets of simplicial complexes that can be analysed by means of  topological theory. The output of the algorithm provides topological metrics characterising the empirical data and  ways to compare different datasets. 

To introduce these ideas more formally, let us consider the canonical example of a set of $N$ points embedded in a metric space, as mentioned in the introduction.  We fix a distance  $\tau$ and say that 2 data points $i$ and $j$  are connected if balls of radius $\tau$ with centres at $i$ and $j$ have a non-empty intersection. Similarly, 3 points form a triangle if the 3 corresponding balls have a  point in common, and so on. This construction is  called a \v Cech complex, a particular type of simplicial complexes for metric spaces \cite{ghrist2014elementary}. A close alternative construction is the Vietoris-Rips complex, where we form a $k$ simplex whenever $k+1$ points are pairwise within distance $\tau$ \cite{Ghrist08}. Clearly, a Vietoris-Rips complex with given $tau$ has more simplices than a \v Cech Complex on the same $N$ points with $\tau / 2$ threshold. There are other ways to build complexes from data, but, for the time being, let us assume that the object has been built.  Formally, one calls {\it simplex} a convex hull of affinely independent points and its {\it faces} are simplices based on subsets of points. Then an {\it abstract simplicial complex} is a set of simplices with the following rules:
\begin {itemize}
\item Every face of a simplex in a complex is in the complex
\item The non-empty intersection of two simplices is a face of
each of them
\end {itemize}
Clearly the \v Cech complex verifies the properties of an abstract simplicial complex. By definition, the dimension of a simplex is equal to the number of its vertices minus one. Note that this quantity should not be confused with the dimension of the original metric space. Thus a single node is a 0-simplex, an edge is a 1-simplex, etc. Let us now take a simplicial complex and the set $K$ of all $k$-simplices, for some fixed some dimension $k$. A {\it $k$-chain} is defined as the formal sum $\sum_i a_i \sigma_i$ for $\sigma_i \in K$ and $a_i$ in some field $F$. A common convention is to take $a_i \in Z_2$, so that the chain  either contains a certain simplex, or not. It is clear that, for a fixed object and dimension, a set of chains is a vector space $C_k$. 

\begin{figure}
\centering
\includegraphics[width=0.5\textwidth]{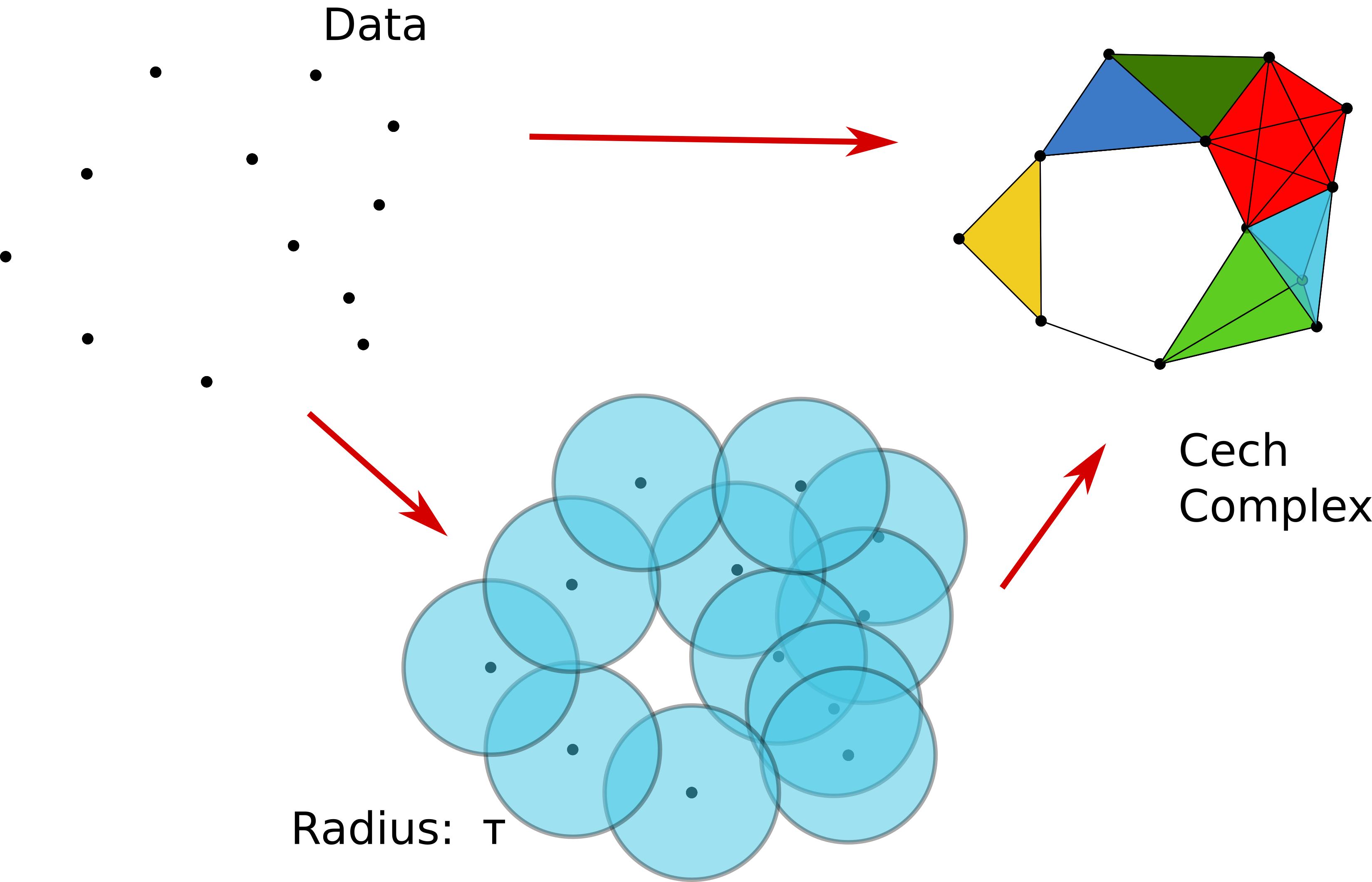}
\caption{Example of a fixed set of points completed to a \v Cech complex with radius $\tau$.}
\end{figure}

Take a simplex $\sigma=[v_0,v_1,...v_k]$ in a complex. By definition, all of its faces also belong to the complex. This is true, in particular, for  faces of dimension $k-1$. This property permits to define a {\it boundary operator} $\delta_k(\sigma) = \sum_i (-1)^i [v_0,...,v_i',...,v_k]$, where $v_i'$ indicates that vertex $v_i$ is deleted from the list. The corresponding transformation defines a linear homomorphism from $C_k$ to $C_{k-1}$. In words the boundary operator maps a simplex to a formal sum of its faces, so for example the boundary of a triangle is simply the sum of its edges. It can be shown that $\delta_{k-1} (\delta_k  (d)) = 0$ for any complex $d$. These concepts allow to define a $k$-chain $c$ ({\it for cycle}) as a simplex verifying $\delta_k (c) = 0$ and a $k$-chain $b$ ({\it for boundary}) if $\exists d \in C_{k+1}$ such that $\delta_{k+1} (d) = b$. Let $B_k$ be the set of all $k$-boundaries and $Z_k$ the set of all $k$-cycles, then we have $B_k \subseteq Z_k \subseteq C_k$ and, due to linearity of $\delta$, they are actually subgroups.

The last paragraph clearly shows that the mathematics of simplicial complexes are far more challenging than those of networks, for instance. For more through introductions on the topic, we point the reader to references such as \cite{Otter17,Patania17_2,sizemore2018importance, chazal2017introduction}. Let us instead focus on the computational  tools that emerge from these formal definitions and what can be learnt from them.
As a first step, we define the {\it k-th homology group} $H_k=Z_k/B_k = \ker \delta_k / $im $\delta_{k+1}$. 
By construction, two simplicial complexes obtained from different datasets are topologically equivalent, and are said to have the same shape, if they produce the same homology groups.
The {\it k-th homology group} directly provides an important, interpretable quantity, as its rank gives the {\em k-th Betti number},  equal to 
the number of k-dimensional holes in the topological fabric.

\begin{figure}
\centering
\includegraphics[width=0.4\textwidth]{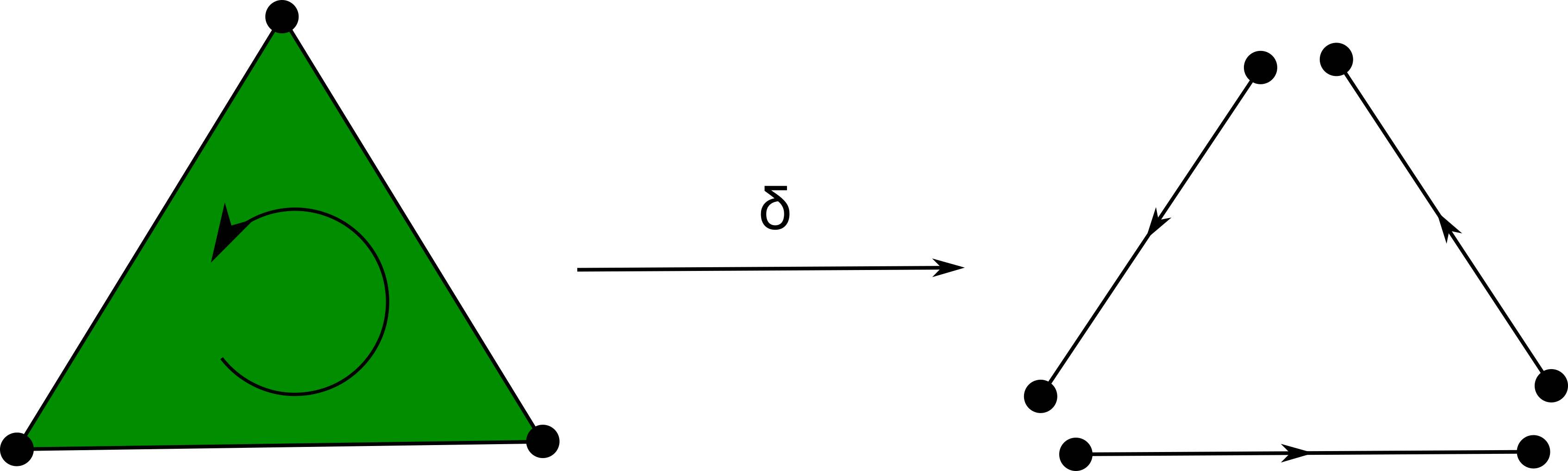}
\caption{The boundary operator of a 2-simplex returns the formal sum of its edges. Arrows represent orientation.}
\end{figure}

In an applied setting, an important aspect is that the construction of the simplicial complex depends on the value of certain parameters, and that the choice of the value can not be done a priori. Take the value of $\tau$ in a \v Cech complex for instance. Increasing values of $\tau$ lead to larger and larger simplices, until a simplex made of the $N$ points is formed, but what value of $\tau$ should be chosen? This will depend on the kind of data and on what is the object of investigation, but clearly the choice will affect the shape of data and the resulting analysis using topological tools.
Persistent homology \cite{Carlsson05,Ghrist08,Edelsbrunner08} is a way to consider the information obtained from all values of parameters, and to represent into an  understandable and easy-to-interpret form. 

For the purpose of this discussion, let us consider again a \v Cech complex for the sake of simplicity. As $\tau$ increases, more and more simplicies are formed in the complex but, as 
 no simplices are deleted, if  a cycle exists for $\tau=\tau_1$, then it will be present for all $\tau > \tau_1$. Along the way,  new cycles can also appear as well as new boundaries. Homology groups thus evolve when $\tau$ changes and persistent homology aims at extracting information from this evolution. 
Tuning the value of $\tau$ for a \v Cech complex is  an example of the general idea of {\it filtrations}. By definition, a {\it filtration} is a nested sequence of simplicial complexes $\emptyset \subseteq F_1 \subseteq F_2 \subseteq ... \subseteq F_n$, which implies a partial order on simplices. By looking at homology groups for different values of the filtration, we  may track when  holes appear (birth) and disappear (death), which can be visualised by means of so-called barcodes. The length of the parameter interval between the death and the birth of a hole is called its {\it persistence}. It is usually assumed that holes with long persistence convey important information about the system, while short ones are associated to noise, but there are cases in which short-lived holes may convey important information \cite{Stolz17}.
Note that the filtration can be performed on other model parameters, for instance on the time when a hole appears or disappears \cite{Salnikov18}, in the case of temporal data, or on the weight of a simplex, in the case of weighted simplices \cite{Petri13}. 
Another important concept is that of persistence landscape \cite{Edelsbrunner02}. Persistence landscapes are piecewise-linear functions, defined in a separable Banach space, that convey the information about births and deaths of homological cycles during the filtration. The great advantage they offer with respect to other visualisations (e.g. barcodes) is that it is possible to compute average landscapes over different datasets and also to define an $L^p$ distance between landscapes, which allows to do statistical analysis on mesoscale structures \cite{Bubenik15}.

\section{From networks to simplicial complexes for complex systems}

The operation leading  to \v Cech complex is clearly reminiscent of the notion of geometric graph \cite{barthelemy2011spatial}. In the latter, a network is constructed from points in a metric space via a threshold, and pairs of nodes are connected by an edge. 
Note  that the relations between networks and geometry are both-ways. Complementary to geometric graphs, aiming at defining networks from geometrical data, many network methods aim at embedding nodes on a metric space. Important examples include hyperbolic embedding methods \cite{krioukov2010hyperbolic}, that naturally produce  
heterogeneous degree distributions and strong clustering in complex networks, where the nodes are assigned a location in an hyperbolic space. The resulting embedding can then be used to help routing information on the network for instance. Spectral embeddings such as diffusion maps \cite{Coifman2005PNAS-1} also play an important role. Related to the notion of kernel on graphs \cite{fouss2007random}, they allow to project networks a lower-dimensional spaces and to define distance or similarity matrices between nodes. The latter can then be exploited to cluster the nodes, e.g. by means of k-means, in order to reveal modules hidden in the original network \cite{schaub2017many}. 

Despite these connections with geometry, most of the emphasis in network science is put on the existence of pairwise edges between nodes in an abstract space, and  on indirect paths of interactions formed by a succession of edges. This focus on connectivity has led to important contributions in our understanding of complex systems, as networks naturally provide a bridge between structure and dynamics \cite{newman2003structure}.
For instance, certain types of interaction networks tend to facilitate diffusion, or to allow for complex dynamical regimes. Important structural properties include the degree distribution, as the existence of high degree nodes provides shorter paths and accelerates diffusive processes \cite{pastor2001epidemic}, the over-representation of motifs that allow for local reinforcement in the dynamics \cite{o2015mathematical}, and community structure, as the presence of dense communities is associated to different time scales for the dynamics and allows for the coexistence of different states \cite{simon1977aggregation}. More importantly,  complex systems are composed by large numbers of interacting elements, and network science thus provides a universal language, applicable to a variety of domains, in order to decipher the myriads of connections in the system.

Despite these successes, network science also has important limitations that may prevent it to properly represent the complexity of  real-world interacting systems. Moreover, these limitations become more and more apparent with the availability of rich datasets, allowing to track paths of diffusion in systems, and the temporal characteristics of the system evolution \cite{lambiotte2016rich}. For these reasons, different attempts are currently developed in order to develop appropriate models for higher-order networks \cite{lambiotte2018understanding}. 
One of those models builds on TDA and represents systems as simplicial complexes. From our previous discussion on the \v Cech complex, it is clear that 
an important difference between networks and simplicial complexes is the possibility to encode higher-order interactions, involving more than 2 nodes. 
Simplicial complexes thus provide an alternative to hypergraphs \cite{ghoshal2009random}, where  many-body interactions can also be encoded but, at the same time, they put a particular emphasis on the underlying the geometrical nature of system. For these reasons, simplicial complexes appear to be a good candidate to model and analyse systems composed of many elements, interacting via many-body interactions, and expected to have a strong geometrical nature. 
When adopting a TDA approach,  it is primordial to remember that the underlying assumption of TDA is that the shape of data matters, and that the existence of a lower-dimensional embedding may correspond to underlying symmetries or constraints in the system. A good example would be noisy observations of a dynamical system in a  periodic orbit. TDA naturally provides tools to detect and quantify such recurrent motion. In that case, the identification of voids in the data points indeed gives us  critical information about the system behaviour. In this direction,  \cite{Maletic16} introduced a method to reconstruct the phase space of a real-world dynamical system from time series using simplicial complexes, which preserve the topology of the state space. In that case, short-lived holes may be interpreted as temporary inaccessible subspaces of the state space, while persistent holes as permanent obstacles to the dynamics.
It is also important to bear in mind the additional conceptual and computational costs associated to TDA, as the number of faces rapidly explodes even in fairly small systems.

Before turning to a presentation of tools and their applications, note that
there exist different ways by which data points can be encoded into simplicial complexes. The first type of encoding deals with points embedded into a metric space, as discussed above, for which standard methods like the \v Cech complex and the Vietoris-Rips complex can be built. There exist also many situations when the system has the structure of a hypergraph, with nodes interacting via non-binary interactions, and where the hypergraph can be modelled as a simplicial complex, called a nerve complex. Potential applications can be found in chemistry, with biochemical reactions may involve more than two species in a reaction \cite{klamt2009hypergraphs}, and social systems, where group interactions naturally appear in collaboration \cite{estrada2005complex} and contact \cite{sekara2016fundamental} networks. In that case, each clique of size $d$ in the hypergraph leads to the a simplex of dimension $d-1$, as well as all its intrinsic simplicies. Finally, simplicial complexes can also be used to analyse standard networks. A first method consist in embedding the nodes in a metric space, as was discussed earlier in this section, and then to construct e.g. a \v Cech complex. For a spectral embedding, the underlying geometry of the problem would thus be associated to the dominant eigenvectors of the network. Alternatively, 
a clique-complex can be constructed by associating each clique in the network to a simplex. 

\begin{figure*}
\centering
\includegraphics[width=0.8\textwidth]{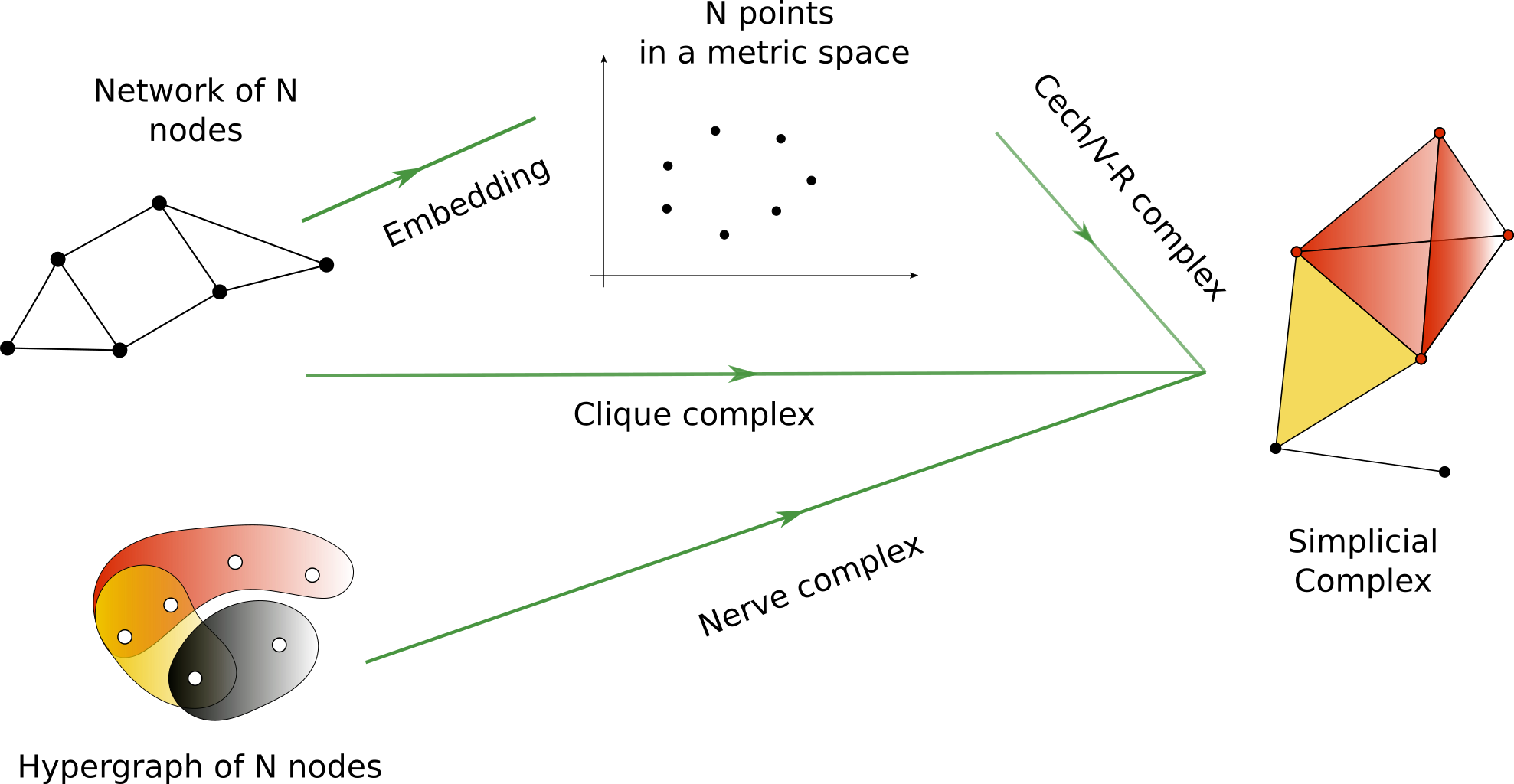}
\caption{A visual summary of the many ways in which simplicial complexes can be built: from a cloud of data points, from a network and from a hypergraph. The simplicial complex represented in the figure corresponds to the hypergraph on its left.}
\end{figure*}

\section {Important tools}

The first algorithm for the computation of persistent homology was provided by \cite{Edelsbrunner02}, which allowed computation over a particular {\it field}, the field $\mathcal{F}_2$, and was later extended for computation over general fields   \cite{CarlssonZomo05}.
Algorithms for computing homology are based on the reduction algorithm, which we briefly describe here following \cite{CarlssonZomo05}. The boundary operator $\delta_k$ can be represented by an integer matrix $M_k$ with entries in $\{-1,0,1\}$, where columns are the $k$ simplices and  rows are the $(k-1)$ simplices. The null space of $M_k$ corresponds to $Z_k$ and the range-space to $B_{k-1}$.
The reduction algorithm uses elementary row and column operations to reduce $M_k$ to its \emph{Smith normal form} $\tilde{M}_k$,

\begin{equation}
\tilde{M}_k =
\left[
\begin{array}{ccc|ccc}
b_1 & & 0 & & & \\
 & \ddots & & & 0 & \\ 
0 & & b_{l_k} & & &  \\
\hline
&  & & & & \\
& 0 & & & 0 &  \\ 
&  & & & &
\end{array}
\right]
\end{equation}

The computation of the Smith normal form in all dimensions is sufficient to characterize the homology of the complex, as $\text{rank} \tilde{M}_k = \text{rank} {M}_k = l_k$, $\text{rank} Z_k = m_k-l_k$ and $\text{rank} \tilde{B}_k = \text{rank} \tilde{M}_{k+1} = l_{k+1}$, so the Betti number can be computed as

\begin{equation}
\beta_k = m_k - l_k- l_{k+1}.
\end{equation}
The vector space  defined above captures the homology of a single complex, while a filtration is constituted by many complexes. To compute persistent homology, one needs to find a basis that is compatible across the entire filtration, which can be shown to exist \cite{CarlssonZomo05}.
A simplified version of the reduction algorithm over a general field is given in  \cite{CarlssonZomo05}. In the worst case, its complexity  is cubic in the number of simplices. Other algorithms have been developed for the reduction of the matrix representation of the boundary operator, and while some of them can be faster for some specific datasets, the complexity remains cubic in general.

Several publicly available software packages  implement reduction and persistence homology: \textsc{javaPlex} \cite{javaplex}, \textsc{Perseus} \cite{perseus}, \textsc{Dionysus} \cite{dionysus}, \textsc{PHAT} \cite{Phat}, \textsc{DIPHA} \cite{dipha}, \textsc{Gudhi} \cite{gudhi}, \textsc{ripser} \cite{ripser} and \textsc{jHoles} \cite{jhole}.  A comparison of their performance \cite{Otter17}, tested  on real and synthetic datasets, concluded that the fastest package overall are \textsc{ripser}, \textsc{Gudhi} and \textsc{DIPHA}, while \textsc{javaPlex} is the easiest and best software for a beginner, as long as  the analysis is done on small complexes.

\section {Application of TDA}

In recent years the number of contributions relying on topological data analysis tools is rapidly increasing, as TDA provides alternative and complementary tools in many disciplines dealing with complex systems. The discipline that is possibly benefiting the most from the adoption of TDA is neuroscience: after the network paradigm proved successful for studying a multitude of problems, the adoption of simplicial complexes and the consequent relaxation of the simplifying assumption that all can be counted for by dyadic relationships seems the natural step forward considering the inherently complex and rich structure of neural systems \cite{Bassett16, Bassett18, Saggar18}.
Several works exploit the fact that a study of homological cycles through persistent homology  provides a way to detect differences between neural systems, in terms of their mesoscale structures. 
For example \cite{Petri14} finds evidence of different brain activity in two states: under the effect of a psychedelic drug and with a placebo. The homological features of brain functional networks in the two cases are significantly different, with an increased integration between cortical region in the altered state.
Developing on the same methodology, \cite{LordPetri} show that measures of edges importance in homological cycles complement the information provided by standard graph metrics. TDA has also proved useful to understand the structure of the brain, \cite{Sizemore18} find that homologycal cycles in structural brain networks link regions of early and late evolutionary origin.
A related stream of works uses persistence landscape distance  to detect changes in functional brain networks during learning of a simple motor task \cite{Stolz17}, and the functional equivalence between imagery and perception \cite{IbanezMarcelo18}, providing empirical evidence that functional equivalence is higher for highly hypnotisable individuals, a scenario that was suggested by behavioural studies but could not be confirmed by means of standard computational tools. This result serves as an example of the ability of TDA to identify significant features that would have remained blurred with standard methods, including networks methods.
Other examples of applications include studies on how topology of brain arteries changes with age \cite{Bendich16}, how the homological features of connectivity for hyperactivity disorder and autism spectrum are different compared to healthy cases \cite{Lee11}, how speech-related brain regions connectivity changes in different scenarios of speech perception \cite{Kim15}, and the topological differences between epileptic and healthy EEG signals  \cite{Piangerelli2018}.

Another area of growing interest is that of machine learning and, more specifically, that of neural networks and deep neural networks. This set of methods have proven to be extremely successful at recognising patterns in noisy data, despite the fact that they take, as an input, single data points or pixels. A promising research direction is to combine TDA with neural networks, by using shapes identified by TDA as input data in the machine learning framework. This technical aspect is far from trivial, as  
 persistent diagrams, for instance, contain a variable number of intervals, while  a structured input of a fixed size is typically required in machine learning. One possible solution is through the use of persistence landscapes which, coupled with convolutional neural networks, leads to so-called called a persistent convolutional neural network model \cite{Liu16}. The key here is the piece-wise linearity of  persistence landscapes that makes  back-propagation  straightforward. A successful application of this methodology is in music classification, where an incorporation of the shape of audio signals helped to  outperform the state of the art methods. As a next step, \cite {Hofer18} proposed   to project persistence diagrams on a fixed-size collection of structure elements, to let the algorithm  decide which information to consider in the training, with excellent results for the classification of the shapes of 2D objects. There exist other ways by which TDA and machine learning can be integrated, as in  \cite{Carlsson18}, where the authors use weights of convolutional networks at different training steps as an input for TDA, and many more ways are expected to come in 
 this
 rapidly evolving field of research.

The applications of TDA in other fields of science are  more sparse and fragmented, but several examples show its growing interest and its potential in a range of problems. Several works aim at classifying weighted networks in terms of the persistence of their homological structures  \cite{Petri13}. This is the case in biology, for instance, where the method  proved successful in identifying high-survival breast cancers \cite{Nicolau11} and in deciphering   noisy biological signals those in electromyographic data  \cite{Phinyomark17}.
 Persistent homology has also been used in genomic datasets to identify evolutionary patterns of RNA viruses \cite{Chan13}.
Applications in finance and economics include the detection of crisis in financial markets \cite{Gidea18}, a description  of  
 the connectivity of the banking networks using Betti numbers \cite{delaConcha2018}, and a classification of countries in terms of the homology of their trade networks 
 \cite{Schauf16}. In computational social science, works have identified patterns of international communities in mobile phone data \cite{Bajardi2015}. Other research areas include dynamical systems, through the study of contagions maps \cite{Taylor15}, and scientometrics, to understand scientific collaborations
\cite{Carstens13, Patania17} and co-occurrences relations of concepts in scientific articles \cite{Salnikov18}.

\section {Perspectives}

The study of networks has emerged as a science through the development of  different facets that feed each other: the design of  algorithms, the identification of  ubiquitous patterns, the identification of mechanism leading to the emergence of these patterns, and an understanding on how they affect the behaviour of the system.  As an example, take the modular organisation of complex networks, which has led to the design of efficient community detection algorithms, a quantification of modularity in a variety of systems, the identification, amongst others, of evolutionary mechanisms driving the formation of communities, and a study of their impact on synchronisation and diffusion \cite{meunier2010modular}. The combination of these findings have helped capture the evasive nature of complex systems. For TDA to achieve a similar status as a building block of the science of complex systems, much remains to be done. As we have discussed, much attention has been dedicated, so far, at the design of computational methods and the study of empirical systems. But large sections remain overall unexplored. A central yet intriguing notion is that of hole. Assuming that a system presents certain properties about the temporal properties and size of its holes, we still lack understanding of the impact of these holes on the function and behaviour of the system. Likewise, where does the complex topology emerge? Can we find simple rules that explain why various datasets from complex systems can be efficiently characterised by means of TDA, as several empirical studies suggest? 

As we have written before, we believe that TDA
is a promising research venue in situations where the salient features of a system are its connectivity and geometrical patterns. It is thus unsurprising that a majority of {\em successes} associated to TDA have been found in neuroscience. Similarly, another promising field of study, still relatively unexplored, would the study of complex urban systems, as cities also present a multi-scale inter-connected organisation and their 
 morphology plays a crucial role in the functioning \cite{batty1994fractal}. In terms of impact on dynamics, finally, interesting research directions include the study of  higher-order analogs of the graph Laplacian,  called the Hodge Laplacian, which could provide mathematical grounds  for the study of diffusion on simplicial complexes \cite{schaub2018}. Together with these considerations, the increasing popularity of TDA is calling for a clarification of the relation between network concepts and simplex geometry \cite{devriendt} and a quantification of the statistical significance of topological features, which motivates models of random simplicial complexes \cite{Kahle2014, Bobrowski2018} and stochastic models of growing complexes \cite{courtney2018dense}.

\bibliographystyle{plain}
\bibliography{rev_biblio}

\end {document}